\documentclass{moriond}
\begin{document}
\title {FINITE TEMPERATURE QCD: QUARKONIUM SURVIVAL AT AND BEYOND $T_c$ AND DIMUON PRODUCTION IN HEAVY ION COLLISIONS}

\author{ALEJANDRO AYALA $^{(1),(2)}$, C. A. DOMINGUEZ $^{(2)}$ \footnote{Speaker}, M. LOEWE $^{(2),(3),(4)}$}
 
\address{$^{(1)}$ Instituto de Ciencias Nucleares, Universidad Autonoma de Mexico, Apartado Postal 70-543, Mexico DF 04510, Mexico}
\address{$^{(2)}$ Centre for Theoretical and Mathematical Physics, 
and Department of Physics, University of
Cape Town, Rondebosch 7700, South Africa}
\address{$^{(3)}$ Instituto de Fisica, Pontificia Universidad Catolica de Chile,
  Casilla 306, Santiago 22, Chile}
\address{$^{(4)}$ Centro Cientifico-Tecnologico de Valparaíso, casilla 110-V, Valparaíso, Chile}

\maketitle\abstracts{
Finite temperature QCD sum rules are applied to the behaviour of charmonium and bottonium states, leading to their survival at and beyond the critical temperature for deconfinement. Di-muon production in heavy-ion collisions in the $\rho$-region is also discussed.}

\section {Thermal QCD Sum Rules}
In this talk I briefly review the method of QCD sum rules (QCDSR), and its extension to finite temperature \cite{Pioneer}. The basic objects in QCDSR are local current-current correlators built from  QCD as well as from hadronic fields, e.g. the correlator of conserved vector currents is
\begin{eqnarray}
\Pi_{\mu\nu} (q^{2})   &=& i \, \int\; d^{4} x \, e^{i q x} \,
<0|T( V_{\mu}(x)   V_{\nu}^{\dagger}(0))|0> \nonumber \\ [.3cm]
&=& (-g_{\mu\nu}\, q^2 + q_\mu q_\nu) \, \Pi(q^2)  \; ,
\end{eqnarray}
where $V_{\mu}(x)$ involves either quark fields or hadronic fields.
Subsequently, a relation between the two representations is obtained by invoking Cauchy's theorem in the complex squared energy $s$-plane, Fig. 1 (Right Panel),  (global quark-hadron duality), i.e. 
\begin{equation}
\int\limits_{0}^{s_0} ds \,f(s)\, \frac{1}{\pi}\, {\mbox{Im}} \Pi(s)= - \frac{1}{2 \pi i} \, \oint_{|s|=s_0} f(s)\, \Pi(s) \, ds \,,
\end{equation}
where $f(s)$ is some analytic function of $s$, the left-hand-side involves the hadronic spectral function, and the right-hand-side the QCD representation.
This allows for a relation between QCD and hadronic physics. After introducing the temperature through quark and hadron propagators, the sum rules lead to results on the temperature behaviour of hadronic and QCD parameters, from $T=0$ up to the critical temperature for deconfinement $T_c$. This is done analytically, in contrast to numerically as in lattice QCD (LQCD).
Among the results are e.g. the near equality of the critical temperatures for chiral-symmetry restoration and quark deconfinement \cite{CHSR}, the confirmation of deconfinement in light-light \cite{rho}  and heavy-light quark bound states \cite{heavy-light}, etc. In contrast, for the case of heavy-heavy quark hadrons the method predicted the survival of charmonium ($J/\psi$, $\eta_c$, $\chi_c$) \cite{charm1}-\cite{charm2}, and bottonium ($\Upsilon$, $\eta_b$) \cite{bottom}.
\begin{figure*}[t]
\begin{center}
\includegraphics[scale=0.30]{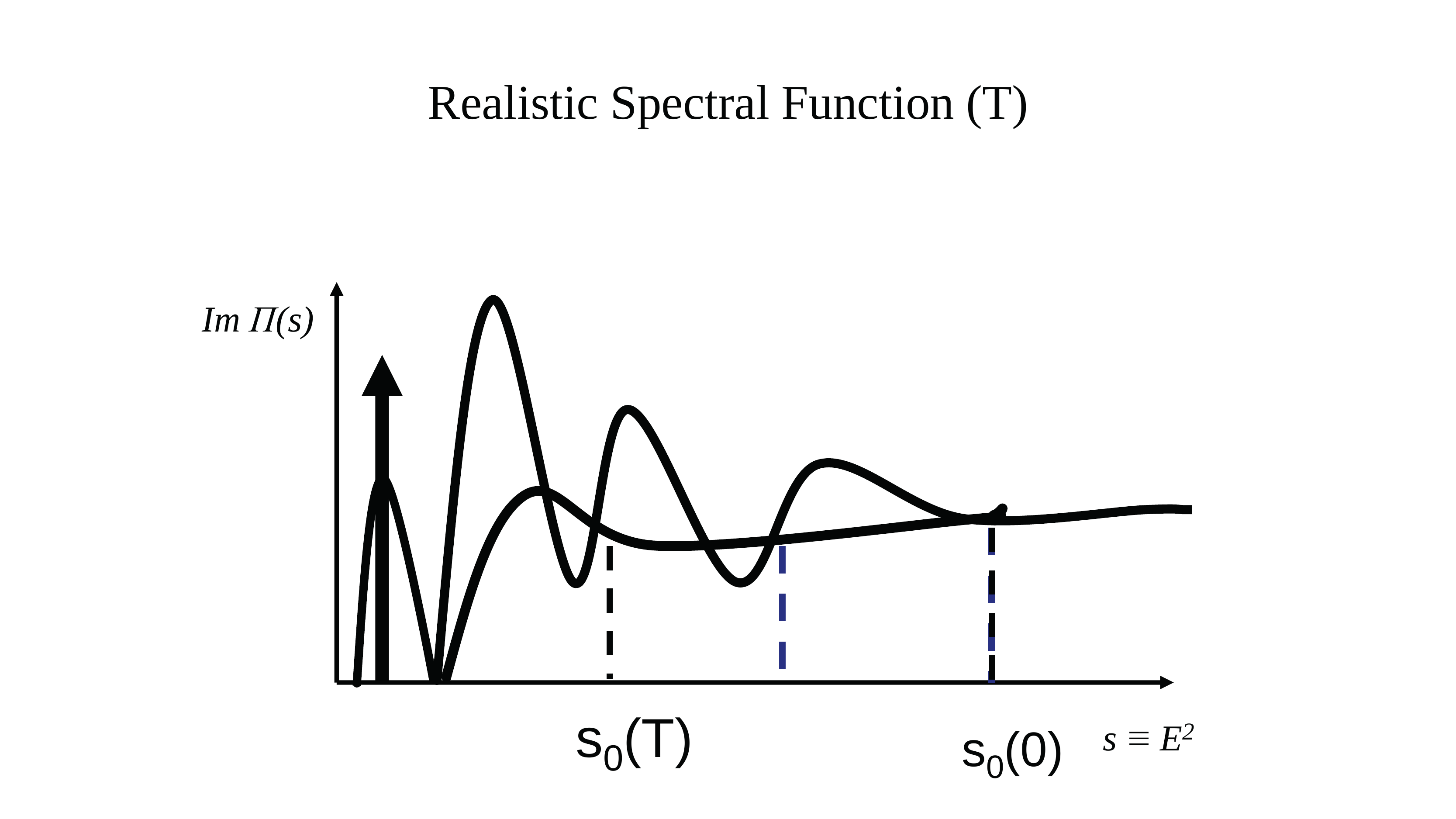}
\includegraphics[scale=0.30]{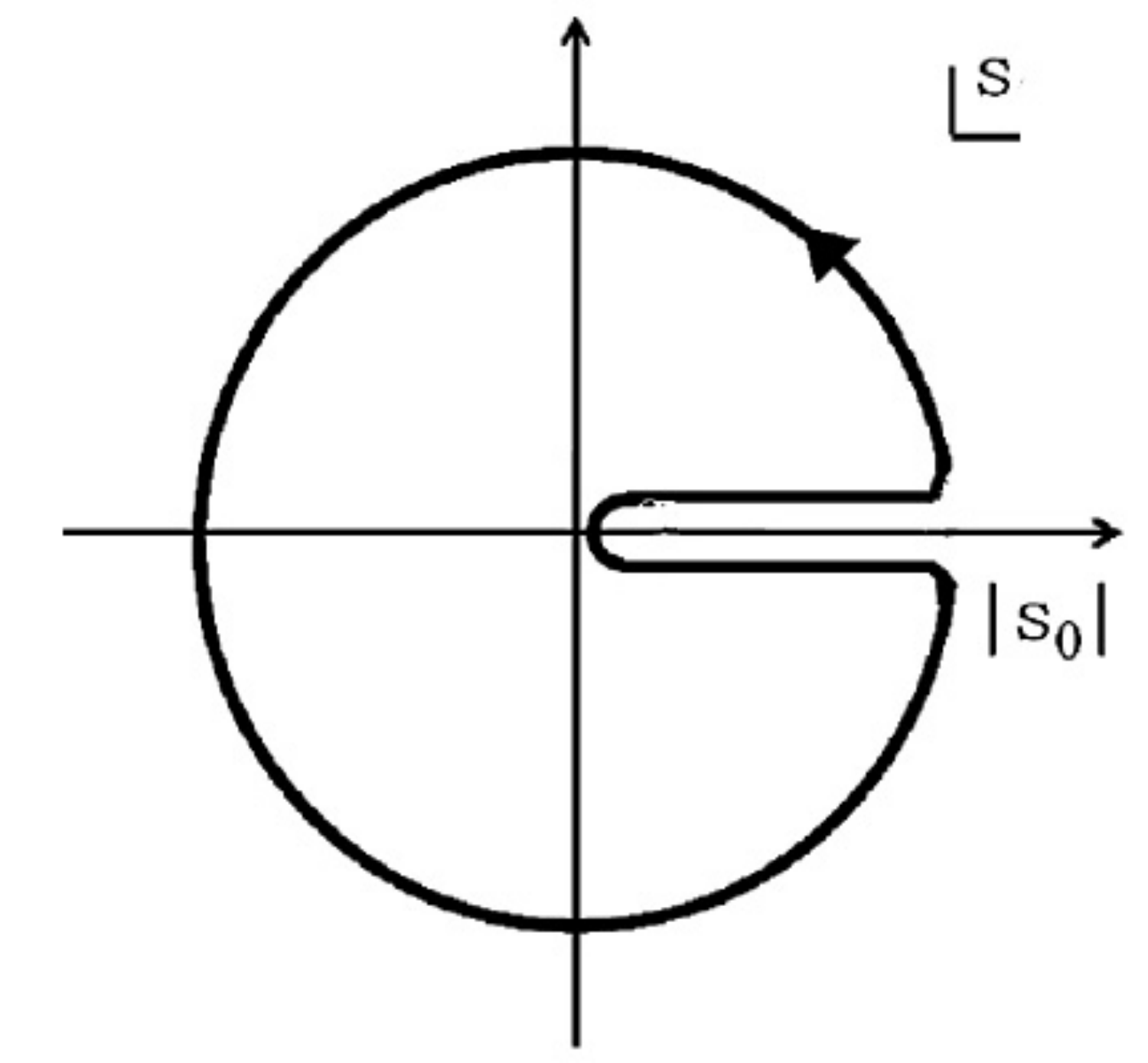}
\end{center}
\caption{\footnotesize{Left: Hadronic spectrum at zero and at finite temperature. Right:  Complex squared-energy s-plane.}}
\label{fig1}
\end{figure*}
\begin{figure*}[b]
\begin{center}
\includegraphics[scale=1.3]{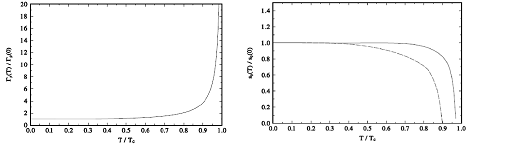}
\end{center}
\caption{\footnotesize{Left: Rho-meson width $\Gamma_\rho(T)/\Gamma_\rho(0)$ as a function of $T/T_c$. Right: The QCD threshold ratio $s_0(T)/s_0(0)$ in the axial-vector channel (dash curve), and in the vector channel (solid curve) as a function of $T/T_c$.}}
\label{fig2}
\end{figure*}
\begin{figure*}
\begin{center}
\includegraphics[scale=1.3]{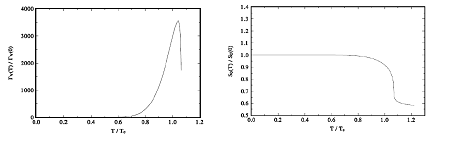}
\end{center}
\caption{\footnotesize{Left: $J/\psi$ width $\Gamma_V(T)/\Gamma_V(0)$ as a function of $T/T_c$. A contrast with Fig.1 (Left Panel) indicates $J/\psi$ survival at and beyond $T_c$. Right: The QCD threshold ratio $s_0(T)/s_0(0)$ as a function of $T/T_c$ for the $J/\psi$. Notice the vertical scale indicating the non-vanishing of this ratio, thus $J/\psi$ survival at and beyond $T_c$ .}}
\label{fig3}
\end{figure*}

In order to introduce $T$-dependent phenomenological parameters, both hadronic as well as QCD, one considers a typical hadronic spectral function depicted in Fig.1 (Left Panel) as a function of the squared energy, $s\equiv E^2$ . At $T=0$ there might be a a pole on the positive real axis, corresponding to a hadronically stable particle, e.g. the pion. This is followed by some resonances (poles on the second-Riemann sheet) with finite widths/lifetimes. Beyond some threshold value  $s = s_0$, the spectral function becomes smooth, and should be accounted for by perturbative QCD (PQCD). There is ample evidence for such a scenario from data on e.g. electron-positron annihilation into hadrons, tau-lepton decay, etc. Turning on the temperature will result in a rearrangement of the spectrum, the higher the temperature the more the distortion. The pole on the positive real axis will move into the second Riemann sheet, thus generating a $T$-dependent finite width for the hadron. At the same time, the original resonances will become broader, while the PQCD threshold $s_0$ will move towards the origin. 
As $T$ approaches the critical temperature for deconfinement, $T_c$, the resonances would have melted and $s_0(T_c)$ would be close to, if not at the origin. In summary, the phenomenological hadronic parameters are expected to be the width, $\Gamma(T)$, and the coupling of the resonance, $f(T)$, to the current entering the correlator $\Pi(s)$. The resonance mass, while potentially dependent on the temperature, it is not a relevant order parameter. In fact, even if the resonance mass would eventually approach the origin, it would not disappear from the spectrum unless its coupling would vanish, and/or its width diverge. In practice it is found that the resonance mass depends only mildly on temperature, in some channels decreasing, and in others increasing slightly with increasing $T$. The phenomenological QCD order parameter is identified with $s_0$, which should be a decreasing function of $T$, vanishing at $T=T_c$ to signal deconfinement.\\
\section*{2 Quarkonium Results}
In order to have reference results Fig.2(Left Panel) shows the thermal behaviour of the rho-meson width \cite{rho}, clearly indicating deconfinement at $T=T_c$. Figure 2(Right Panel) shows the behaviour of the QCD threshold $s_0(T)/s_0(0)$ in the axial-vector (pion, $a_1$) channel (dash curve) and in the vector channel (solid curve). The former corresponds to chiral-symmetry restoration, and the latter to deconfinement. The slight difference in critical temperatures is well within the uncertainty of the method. As expected, the QCD threshold moves towards the origin, as the width diverges. Not shown is the behaviour of the current-hadron coupling (the $\rho$ and the $a_1$ mesons), which decreases monotonically with increasing $T$, vanishing at $T=T_c$. Qualitatively similar results are obtained in the case of heavy-light pseudoscalar and vector mesons \cite{heavy-light}.
In contrast to these results, the situation in the heavy-heavy-quark sector is entirely different \cite{charm1}-\cite{bottom}. In fact, while the width initially increases, as for light quark systems, it eventually decreases substantially at and above $T_c$ (see Fig.3 (Left Panel)). The QCD threshold, $s_0(T)$ while decreasing with increasing $T$, it does so only mildly  up to 40\%, and remains non-zero and frozen well above $T_c$ (see Fig.3(Right Panel)). Finally, not shown here, the current-hadron coupling initially remains fairly constant, to eventually {\bf increase} substantially at and above $T_c$ (see Fig.4 in \cite{charm1}). Similar results are obtained in the scalar and pseudoscalar charmonium channels \cite{charm2},
thus unambiguously pointing to the survival of charmonium. Finally, results for bottonium in the vector ($\Upsilon$) and pseudoscalar ($\eta_b$) channels \cite{bottom} are similar to those of charmonium, except that solutions to the QCD sum rules do not extend beyond $T_c$. It is rather important to mention that lattice QCD has obtained, for the first time, results  for the thermal widths of bottonium \cite{LQCD}, in full qualitative agreement with \cite{bottom}.
It is known that potential models fail to obtain these results. It is likely that this is due to the absence in these models of a centre cut in the complex energy $\omega$-plane ($s\equiv \omega^2$), extending from $\omega = -|\bf {q}|$ to  $\omega = +|\bf {q}|$, with $\bf{q}$  the three-momentum. This is a purely relativistic effect, usually referred to as the {\it scattering term}, present in some channels, and absent in others\cite{Pioneer}. When a reference frame at rest with respect to the thermal medium is chosen (vanishing three-momentum), the centre cut in the energy-plane shrinks to a point, leading in some cases to the appearance of a delta function of the energy, and to a null result in others. This term depends quadratically on the temperature, thus contributing mainly close to the critical temperature. It also may have a dependence on the current quark mass associated with the correlation function. For light quarks the scattering term is independent of the quark mass, but for heavy quark correlators the scattering term is proportional to the square of the quark mass, thus rendering it important  close to $T_c$.
\begin{figure*}
\begin{center}
\includegraphics[scale=0.5]{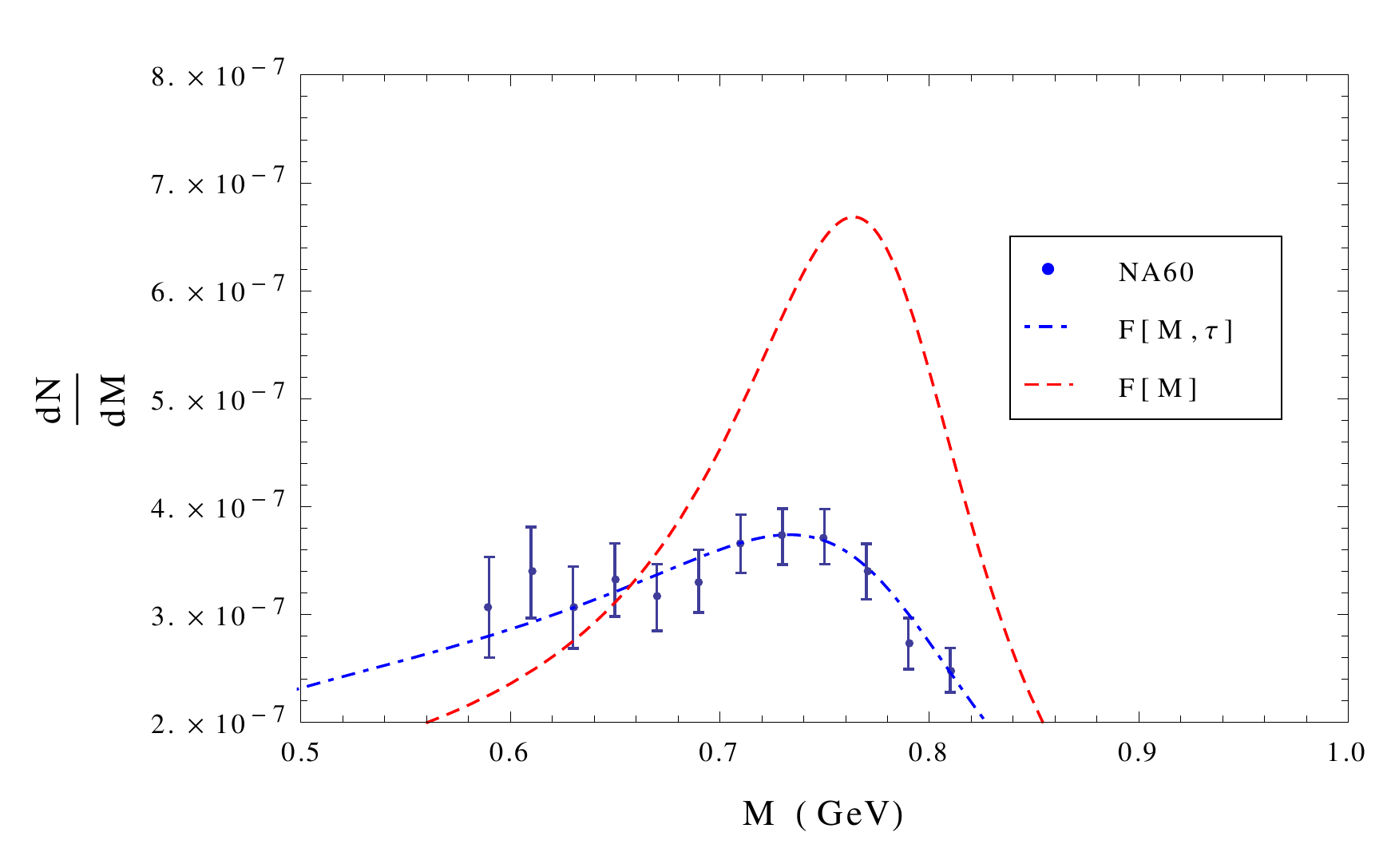}
\end{center}
\caption{\footnotesize{Dimuon production in In-In collisions from the NA60 Collaboration. Dash-dot line is the QCD prediction with independently determined $\rho$-meson thermal parameters. Dash curve is for no thermal dependence.}}
\label{fig4}
\end{figure*}
\section*{3 Dimuon Production in Heavy-Ion Collisions}
Once the $T$-dependent $\rho$-meson parameters, mass, width, and coupling, $M_\rho$, $\Gamma_\rho$, and $f_\rho$ , have been determined \cite{rho} it is straightforward to predict the di-muon production rate for In-In collisions, as measured at CERN by the NA60 collaboration \cite{NA60}. This  parameter-free prediction \cite{dimuon} is shown in Fig. 4 (dash-dot line) together with the NA60 data, and the rate for the case of $T$-independent $\rho$-meson parameters (dash curve). The $\rho$-meson broadening effect, first proposed in this context some 25 years ago \cite{dimuon0}, is quite dramatic. It should be emphasized, though, that this QCD result is not expected to account for the di-muon spectrum away from the $\rho$-meson region. To accomplish this one would need models of hadronic interactions \cite{Rapp}.
\section*{4 Acknowledgments}
The speaker thanks the organizers for the invitation and the opportunity to deliver this talk. He also thanks his various collaborators in the projects being referred to.

\end{document}